\documentclass[11pt]{amsart}
\usepackage{amsmath}
\usepackage{amssymb}
\usepackage[latin1]{inputenc}
\usepackage[T1]{fontenc}

\newcommand{\be}{\begin{equation}}
\newcommand{\eeq}{\end{equation}}
\newcommand{\bet}{\begin{equation*}}
\newcommand{\eeqt}{\end{equation*}}
\newcommand{\bea}{\begin{eqnarray}}
\newcommand{\eeqa}{\end{eqnarray}}
\newcommand{\beat}{\begin{eqnarray*}}
\newcommand{\eeqat}{\end{eqnarray*}}
\newcommand{\goesto}{\longrightarrow}
\newcommand{\h}[1]{\mathcal{#1}}
\newcommand{\hil}{\mathcal{H}}

\newcommand{\hD}{\mathcal{D}}

\newcommand{\hA}{\mathcal{A}}
\newcommand{\hB}{\mathcal{B}}

\newcommand{\piker}{\frac {1}{2\pi}}

\newcommand{\C}{\mathbb{C}}
\newcommand{\N}{\mathbb{N}}

\newcommand{\R}{\mathbb{R}}

\newcommand{\RII}{\mathbb{R}^2}

\newcommand{\LkR}{L^2(\R)}
\newcommand{\bra}{\langle}
\newcommand{\ket}{\rangle}
\newcommand{\vp}{\varphi}

\newcommand{\qker}{\frac {1}{\sqrt{2}}}

\newcommand{\cc}[1]{\overline{#1}}

\newcommand{\binc}[2]{{{#1} \choose {#2}}}

\newcommand{\intL}[2]{L(#1,#2)}
\newcommand{\intd}[2]{D(#1,#2)}
\newcommand{\intsqd}[2]{\widetilde{D}(#1,#2)}

\newcommand{\LEnx}{\intL{x^k}{E^{|n\ket}}}
\newcommand{\LEny}{\intL{y^k}{E^{|n\ket}}}

\newtheorem{lemma}{Lemma}
\newtheorem{theorem}{Theorem}
\newtheorem{proposition}{Proposition}

\begin{document}
\title[Moment operators]{Moment operators of the Cartesian margins of the phase space observables}
\author{J. Kiukas}
\address{Jukka Kiukas,
Department of Physics, University of Turku,
FIN-20014 Turku, Finland}
\email{jukka.kiukas@utu.fi}
\author{P. Lahti}
\address{Pekka Lahti,
Department of Physics, University of Turku,
FIN-20014 Turku, Finland}
\email{pekka.lahti@utu.fi}
\author{K. Ylinen}
\address{Kari Ylinen,
Department of Mathematics, University of Turku,
FIN-20014 Turku, Finland}
\email{kari.ylinen@utu.fi}
\begin{abstract}
The theory of operator integrals is used to determine the moment operators of the Cartesian margins of the phase space
observables generated by the mixtures of the number states. The moments
of the $x$-margin are polynomials of the position operator and those of the $y$-margin are polynomials of
the momentum operator.
\end{abstract}
\maketitle

\section{Introduction}

According to the theory developed in \cite{Lahti}, each complex valued measurable function $f$ and operator measure $E$
determine a unique, possibly unbounded, linear operator $\intL{f}{E}$, the operator integral of $f$ with respect to $E$. 
In the case of real valued functions and phase space operator measures, a natural application of this theory
is quantization.

In general, quantization means any procedure which associates a quantum mechanical observable to a given
classical dynamical variable, the latter being represented by a real valued measurable function on the phase space
$\R^{2n}$ of the classical system. 
Phase space quantization schemes are often realized by associating to a given classical variable $f$ the operator
$\int f(q,p)\Delta(q,p) dqdp$, where $\Delta$ is some operator valued phase space function and the integral is
interpreted e.g. in the weak or strong sense on a suitable domain  (cf. e.g. \cite{Dubin}, \cite{Landsman}, and
\cite{Schroeck}).

Consider the quantization determined by the operator density $(q,p)\mapsto \Delta(q,p) := (2\pi)^{-1}W(-q,p)TW(-q,p)^*$,
where the $W(-q,p)$ are the Weyl operators acting in a separable Hilbert space and $T$ is a state, i.e. a
positive operator of trace one.
Now the map $\h B(\R^{2n})\ni B\mapsto E(B):=\int_B \Delta\in L(\hil)$ is a phase space observable, so that
any classical variable $f$ can be integrated with respect to it. On its natural domain, the operator integral $\intL{f}{E}$
coincides with the (weak) quantization integral $\int f \Delta$.
In this way, $\intL{f}{E}$ can be interpreted as a quantization of the classical variable $f$.
It can be noted that this approach differs from the Weyl
quantization, which is obtained by replacing $T$ in the above density $\Delta(q,p)$ by (a constant times) the parity operator in
$\LkR$ (cf. e.g. \cite[Sect. IV.1]{OQP}, \cite[p. 199]{Dubin}, and \cite[pp. 140-141]{Landsman}).

In this paper, we consider the phase space observables on $\R^2$ associated with certain pure states and their
convex combinations, with
the state vectors being taken from a fixed countable orthonormal basis of the separable Hilbert space. The moment
operators of the Cartesian margins of these operator measures will be determined using the theory of operator integrals.
The results sharpen and extend some of those obtained previously in \cite{LahtiII}.

section{Operator integrals and phase space observables}

\subsection{The operator integral}
In the following, we review the basic results of \cite{Lahti} on the theory of operator integrals and prove
a proposition concerning integration with respect to sequences of positive operator measures.
 
Let $\hil$ be a Hilbert space, with inner product $\bra \cdot |\cdot\ket$, and $L(\hil)$ the set of bounded operators on
$\hil$. Let $\Omega$ be a nonempty set, $\h A$ a
$\sigma$-algebra of subsets of $\Omega$, and $E:\h A\to L(\hil)$ a positive operator measure
(a positive operator valued set function $\sigma$-additive with respect to the strong, or,
equivalently, the weak operator topology). For all $\vp, \psi\in\hil$, the map $\hA\ni X\mapsto\bra \psi |E(X)\vp\ket\in\C$
is a complex measure, and it is denoted by $E_{\psi, \vp}$. Let $f:\Omega\to\C$ be an $\h A$-measurable function
and let $\intd{f}{E}$ be the set of those vectors $\vp\in\hil$ for which $f$ is integrable with respect to the complex measure
$E_{\psi, \vp}$ for all $\psi\in\hil$. The set $\intd{f}{E}$ is a vector subspace of $\hil$, and there is a unique linear
operator $\intL{f}{E}$ on the domain $\intd{f}{E}$ such that
\bet
\bra \psi |\intL{f}{E}\vp\ket = \int f dE_{\psi, \vp}
\eeqt
for all $\vp\in \intd{f}{E}$ and $\psi\in\hil$ (cf. \cite{Lahti}). We call $\intL{f}{E}$ the \emph{(operator) integral} of $f$
with respect to $E$.

Let $\intsqd{f}{E}$ be the set of those vectors $\vp\in\hil$ for which $|f|^2$ is integrable with respect to the positive
measure $E_{\vp,\vp}$. We have the following results, proved in \cite{Lahti}:
\begin{theorem}
\begin{itemize}
\item[(a)] $\intsqd{f}{E}$ is a vector subspace of $\intd{f}{E}$.
\item[(b)] If $E(X)$ is a projection for all $X\in\hA$, then $\intsqd{f}{E} = \intd{f}{E}$.
\end{itemize}
\end{theorem}
It is well known that, in case (b), the domain is dense.
\begin{theorem}
If $f$ is real valued, then $\intL{f}{E}$ is a symmetric operator, that is,
$\bra \psi |\intL{f}{E}\vp\ket = \bra \intL{f}{E}\psi |\vp\ket$ for all $\psi,\vp\in \intd{f}{E}$.
\end{theorem}

The following general lemma will be used in the proof of Proposition \ref{LsubLnlemma}.

\begin{lemma}\label{measurelemma}
Let $\mu_n:\h A\to\C$ be a complex measure for each $n\in\N$ such that the series $\sum_{n\in\N} \mu_n$
converges absolutely in the total variation norm. Let $\mu$ and $\nu$ denote the measures
$\sum_{n\in\N} \mu_n$
and $\sum_{n\in\N} |\mu_n|$, respectively. Here $|\cdot |$ stands for the total variation.
\begin{itemize}
\item[(a)] $f$ is $\nu$-integrable, if and only if $\sum_{n=1}^\infty \int |f|d|\mu_n| < \infty$.
\item[(b)] If $f$ is $\nu$-integrable, then $f$ is integrable with respect to $\mu$ and all the measures $\mu_n$, and
\bet
\int f d\mu = \sum_{n=1}^\infty \int f d\mu_n.
\eeqt
\end{itemize}
\end{lemma}
\noindent {\bf Proof. } (a) Assume that $f$ is $\nu$-integrable (i.e. $|f|$ is such). Since $|\mu_n(B)|\leq \nu(B)$ for all $B\in\hA$
and $n\in\N$, $|f|$ is $\mu_n$-integrable for each $n\in\N$. Now
\bet
\sum_{n=1}^k \int |f|d|\mu_n| =  \int |f| d\left(\sum_{n=1}^k |\mu_n|\right) \leq\int |f| d\nu
\eeqt
for each $k\in\N$, so that $\sum_{n=1}^\infty \int |f|d|\mu_n| < \infty$.
The converse follows from Lemma A.5. of \cite{Lahti}.

\noindent (b) Clearly $\mu$ and all the $\mu_n$ are $\nu$-continuous.
Let $g$ and $g_n$ be their Radon-Nikod\'ym derivatives
with respect to $\nu$, respectively.
Since
\bet
\left(\sum_{n=1}^k \mu_n-\mu\right)(B) = \int_B \left(\sum_{n=1}^k g_n-g\right) d\nu
\eeqt
for all $B\in\h A$ and $k\in\N$, we have
\bet
\lim_{k\goesto \infty}\int \left|\sum_{n=1}^k g_n-g\right| d\nu = \lim_{k\goesto \infty}\left\|\sum_{n=1}^k \mu_n-\mu\right\| = 0,
\eeqt
where $\|\cdot\|$ denotes the total variation norm.
This means that the series $\sum_n g_n$ converges in $L^1(\nu)$ to the function $g$. Thus some subsequence
$\left(\sum_{n=1}^{k_m} g_n\right)$ converges to $g$ $\nu$-almost everywhere. In addition,
the monotone convergence theorem gives
\bet
\nu(B) = \sum_{n=1}^\infty \int_B |g_n|d\nu = \int_B\left(\sum_n |g_n|\right) d\nu
\eeqt
for all $B\in \h A$, so that
\be\label{ineq}
\left|f(x)\sum_{n=1}^{k_m} g_n(x)\right|\leq |f(x)|\sum_n |g_n| = |f(x)|
\eeq
for $\nu$-almost all $x\in\Omega$.

Assume now that $f$ is $\nu$-integrable. Then by (a), $f$ is $\mu_n$-integrable for each $n\in\N$, and
the series $\sum_{n=1}^\infty \int f d\mu_n$ converges absolutely. Because of (\ref{ineq}), the dominated convergence
theorem implies that $fg$ is $\nu$-integrable, i.e. $f$ is $\mu$-integrable, and
\bet
\int f d\mu = \int fg d\nu = \lim_{m\goesto\infty} \sum_{n=1}^{k_m} \int fg_n d\nu = \sum_{n=1}^\infty \int f d\mu_n.
\eeqt
$\Box$
\begin{proposition}\label{LsubLnlemma}
Let $E^n:\h A\to L(\hil)$ be a positive operator measure for each $n\in\N$ such that the range of $E^n$ is bounded
in norm by $M_n>0$, with $\sum_n M_n <\infty$. Then the norm limit $\sum_n E^n(B)$, $B\in\h A$, defines a positive
operator measure $E$, for which
\bet
\intL{f}{E}|_{\intsqd{f}{E}}\subset \sum_n \intL{f}{E^n}|_{\intsqd{f}{E^n}},
\eeqt
where the series is understood to converge in the weak sense to an operator, whose domain
consists of those vectors $\vp\in \bigcap_{n\in \N} \intsqd{f}{E^n}$ for which the vector series
$\sum_n \intL{f}{E^n}\vp$ converges weakly.
\end{proposition}
\noindent {\bf Proof.} Let $\psi, \vp\in\hil$. Since $\|E^n_{\psi, \vp}\|\leq 4 \sup_{B\in\h A} |E^n_{\psi, \vp}(B)|\leq 4\|\psi\|\|\vp\|M_n$,
the series $\sum_n E^n_{\psi, \vp}$ converges absolutely in the total variation norm. The above inequality also implies
that for every $B\in \h A$, the sesquilinear functional $(\vp,\psi)\mapsto \sum_n E^n_{\psi, \vp}(B)$ is bounded, so that
there is a (clearly positive) operator $E(B)\in L(\hil)$, for which $\bra \psi |E(B)\vp\ket= \sum_n E^n_{\psi, \vp}(B)$.
Thus the map $B\mapsto E(B)$ is a positive operator measure, for which $E(B) = \sum_n E^n(B)$ in the operator norm
and $E_{\psi, \vp}=\sum_n E^n_{\psi, \vp}$ in the total variation norm.

By applying Lemma \ref{measurelemma} (b) to the finite positive measures $E^n_{\vp,\vp}$ and the function $|f|^2$, we
see that if $\vp\in\intsqd{f}{E}$, then $\vp\in\intsqd{f}{E^n}$ for all $n\in\N$, and
$\sum_n\int |f|^2 dE_{\vp,\vp}^n= \int |f|^2dE_{\vp,\vp} < \infty$. Now let $F^n$ denote the positive operator measure
$M_n^{-1}E^n$. We have
$\sum_n M_n \sqrt{\int |f|^2 d F^n_{\psi,\vp}}\leq \sum_n M_n (1+\int |f|^2 d F^n_{\psi,\vp} )<\infty$.
Using the inequality
\bet
\int |f| d|F_{\psi, \vp}^n|\leq \|\psi\|\sqrt{\int |f|^2 dF_{\vp,\vp}^n}
\eeqt
from \cite{Lahti}, we get $\sum_n \int |f| d|E_{\psi, \vp}^n|= \sum_n M_n \int |f|d|F_{\psi,\vp}^n|<\infty$, from which it follows
by lemma \ref{measurelemma} (a) and (b) that
\bet
\bra \psi |\intL{f}{E}\vp \ket = \sum_n \bra \psi |\intL{f}{E^n}\vp\ket.
\eeqt
$\Box$

\subsection{Phase space observables}
We assume that the Hilbert space $\hil$ is separable. For any $\vp,\psi\in\hil$, let $|\vp\ket\bra\psi |$ denote the operator
$\xi\mapsto \bra\psi|\xi\ket\vp$. Let $\{|n\ket\mid n\geq 0\}$ be a fixed orthonormal basis of $\hil$. We call it the number
basis. Let $A_+=\sum_{n\geq 0} \sqrt{n+1}|n+1\ket\bra n|$ and $A_-=\sum_{n\geq 0} \sqrt{n+1}|n\ket\bra n+1|$ be the
raising and lowering operators associated with this basis. They are closed operators with the domain
\bet
\hD(A_+)=\hD(A_-) = \left\{ \vp\in\hil\Big|\sum_{k}k|\bra \vp|k\ket|^2<\infty\right\},
\eeqt
and they satisfy the relation $A_+=A_-^*$. The symmetric operators $\qker(A_++A_-)$ and $\qker i(A_+-A_-)$
are essentially selfadjoint on $\hD(A_+)$ and their closures $Q$ and $P$ are unitarily equivalent to
the canonical position and momentum operators, respectively, acting in $\LkR$. The operators $A_+$ and $A_-$ can
be expressed in terms of $Q$ and $P$ according to
\bea
A_+&=&\qker(Q-iP),\nonumber\\
A_-&=&\qker(Q+iP)\label{AQP}
\eeqa
(cf. \cite[p. 283]{Birman} and \cite[p. 73]{Putnam}). The number operator
$N := \sum_{n\geq 0} n|n\ket\bra n|=A_+A_-$, with the domain
\bet
D(N) = \left\{ \vp\in\hil\Big|\sum_{k}k^2|\bra \vp|k\ket|^2<\infty\right\},
\eeqt
is selfadjoint and satisfies
\be\label{numberop}
N +\frac 12 I = \frac 12 (Q^2+P^2).
\eeq
(The last equality is a simple consequence of (\ref{AQP}) and the fact that $D(N)=D(A_+A_-)=D(A_-A_+)$.)

The operators $Q$ and $P$ generate strongly continuous one parameter unitary groups
$p\mapsto V(p) := e^{ipQ}$
and $q\mapsto U(q) := e^{iqP}$ which satisfy the Weyl relation $U(q)V(p) = e^{iqp}V(p)U(q)$ for all $q,p\in \R$.
The Weyl operators $W(q,p)$ are defined by  $W(q,p) = e^{-\frac 12 iqp} U(q)V(p)$ for $q,p\in\R$. The map
$(q,p)\mapsto W(-q,p)$ is a projective representation of $\R^2$.

Let $T$ be a state operator.
Then
\bet
I=\piker\int_{\RII}W(-q,p)TW(-q,p)^*dqdp,
\eeqt
and the map $E^T:\h B(\RII)\to L(\h H)$ defined by
\bet
E^T(B)=\piker\int_{B} W(-q,p)TW(-q,p)^* dqdp
\eeqt
is a normalized positive operator measure, a phase space observable. Here $\h B (\RII)$ denotes the
$\sigma$-algebra of Borel subsets of $\RII$ and the integrals
are understood in the weak sense. The construction of the operator measures $E^T$ can be found for instance in
\cite{Davies} or \cite{Stulpe}. The operator measure $E^T$ is covariant with respect to the projective representation
$(q,p)\mapsto W(-q,p)$ in the sense that $E^T(B+(q_0,p_0))=W(-q_0,p_0)E^T(B)W(-q_0,p_0)^*$ for all $B\in\hB(\RII)$ and
$(q,p)\in\RII$.

The following characterization is obtained in \cite{CDeV} and \cite{Werner}: every normalized positive operator measure
$E:\hB(\RII)\to L(\hil)$, which is covariant with respect to the representation $(q,p)\mapsto W(-q,p)$, is of the form
$E^T$ for some state $T$. 

We let $E^{|s\ket}$ denote the phase space observable associated with the number state $|s\ket\bra s|$.
Consider the mixed states
\be\label{mixed}
T=\sum_{n=0}^{\infty} w_n |n\ket\bra n|,
\eeq
where $w_n\geq 0$ and $\sum w_n = 1$. These states are the ones for which the observable $E^T$ is covariant
also with respect to the phase shifts in the sense that
\bet
e^{i\theta N}E^T([0,\infty)\times X)e^{-i\theta N} = E^T([0,\infty)\times(X+\theta))
\eeqt
for all $\theta\in [0,2\pi)$ and $X\in\h B([0,2\pi))$, where $\RII = [0,\infty)\times [0,2\pi)$ and the sum $X+\theta$ is
understood modulo $2\pi$. (cf. \cite{Pellonpää}).

Since $w_n\|E^{|n\ket} (B)\|\leq w_n$ for all $n\in\N$ and $B\in\h B(\RII)$,
Proposition \ref{LsubLnlemma} can be applied to the positive operator measures $w_nE^{|n\ket}$. That the norm limit
$\sum_n w_n E^{|n\ket}(B)$ equals $E^T(B)$, follows from the identity
%
\bet
\bra \vp | E^T(B)\vp\ket = \piker \int_B (\sum_n w_n |\bra \vp |W(-q,p)|n\ket |^2) dqdp
= \sum_n w_n \bra \vp | E^{|n\ket}(B)\vp\ket
\eeqt
where $\vp\in\hil$ is arbitrary and the monotone convergence theorem has been used.

\section{Moment operators of the Cartesian margins of the phase space observables associated with the number states}\label{sec3}

Let $x$ and $y$ denote the functions $(q,p)\mapsto q$ and $(q,p)\mapsto p$, respectively.
In \cite{Lahti}, the moment operators $\intL{(x\pm iy)^k}{E^{|n\ket}}$ and
$\intL{(x^2+y^2)^k}{E^{|n\ket}}$ were determined. In \cite{LahtiII}, these results were used to obtain the
operator relations
\bea
\intL{x}{E^{|n\ket}} &\subset& Q,\nonumber\\
\intL{y}{E^{|n\ket}} &\subset& P,\nonumber\\
\intL{x^2}{E^{|n\ket}} &\subset& (n+\frac 12)I+Q^2,\nonumber\\
\intL{y^2}{E^{|n\ket}} &\subset& (n+\frac 12)I+P^2. \label{moments0}
\eeqa
In this section, we determine directly the moment operators $\intL{x^k}{E^T}$ and $\intL{y^k}{E^T}$, where the state $T$
is taken to be of the form $\sum_n w_n |n\ket\bra n|$. The results show, among other things, that the inclusions
(\ref{moments0}) are in fact equalities.

\subsection{The operators $\intL{x^k}{E^{|n\ket}}$ and $\intL{y^k}{E^{|n\ket}}$}

Let $U:L^2(\R)\to \hil$ be the unitary operator which maps the Hermite function basis $\{ h_n \}_{n\geq 0}$
of $\LkR$ onto the number basis of $\hil$ according to the rule $Uh_n=|n\ket$. The position and
momentum operators in $\LkR$ correspond to the operators $Q$ and $P$ via the unitary transformation $U$, so that
the operators $W_0(q,p)$, defined by $W_0(q,p) = U^{-1}W(q,p)U$, act in $\LkR$ according to the formula
\bet
(W_0(q,p)f)(t) = e^{i\frac 12 qp}e^{ipt}f(t+q).
\eeqt
We need the following well-known result (see e.g. \cite[pp. 47 and 49]{Stulpe} ).
\begin{lemma}\label{stulpelemma}
Let $F: \LkR\to\LkR$ denote the Fourier-Plancherel operator. Let $u\in\LkR$ be a unit vector and $f\in \LkR$.
Then $\cc{u}(\cdot-q)f\in L^1(\R)\cap\LkR$ for
almost all $q\in\R$ and $\cc{Fu}(\cdot-p)Ff\in L^1(\R)\cap\LkR$ for almost all $p\in\R$. In addition,
\bet
\frac {1}{\sqrt{2\pi}}\bra W_0(-q,p)u|f\ket = e^{i\frac 12qp}F(\cc{u}(\cdot-q)f)(p)
\eeqt
for almost all $q\in\R$ and all $p\in\R$, and
\bet
\frac {1}{\sqrt{2\pi}}\bra W_0(-q,p)u|f\ket = e^{-i\frac 12qp}F^{-1}(\cc{Fu}(\cdot-p)Ff)(q)
\eeqt
for almost all $p\in\R$ and all $q\in\R$.
\end{lemma}
\noindent {\bf Proof.} Since $u, f\in\LkR$, the function $\cc{u}(\cdot-q)f$ is integrable. Because
$\|f\|^2 = \int |\cc{u}(t-q)f(t)|^2dtdq$ by Fubini's theorem, $\cc{u}(\cdot-q)f\in \LkR$ for almost all $q$. Similarly,
$\cc{Fu}(\cdot-p)Ff\in L^1(\R)\cap\LkR$ for almost all $p\in\R$. The rest of the proof follows from straightforward
calculations. $\Box$

\

First we determine the square integrability domains corresponding to the functions $x^k$ and $y^k$.
\begin{lemma}\label{momentdomainlemma}
$\intsqd{x^k}{E^{|n\ket}}=D(Q^k)$ and $\intsqd{y^k}{E^{|n\ket}}=D(P^k)$ for all $k\in\N$.
\end{lemma}
\noindent {\bf Proof.} Let $k\in\N$ be fixed. Let $\vp\in\hil$ and $f=U^{-1}\vp\in \LkR$. If $\vp\in \intsqd{x^k}{E^{|n\ket}}$,
the function $(q,p)\mapsto q^{2k}|\bra f|W_0(-q,p)h_n\ket|^2 = q^{2k}|\bra \vp|W(-q,p)|n\ket|^2$ is
integrable over $\RII$, and
\beat
\int_{\RII} q^{2k} dE^{|n\ket}_{\vp,\vp}(q,p)
&=& \piker \int q^{2k}\left(\int|\bra \vp|W(-q,p)|n\ket|^2dp\right)dq\\
&=& \int q^{2k}\left(\int|F(\cc{h_n}(\cdot-q)f)(p)|^2dp\right)dq\\
&=& \int q^{2k}\left(\int|h_n(t-q)|^2 |f(t)|^2dt\right)dq\\
&=& \int\left(\int q^{2k}|h_n(t-q)|^2 |f(t)|^2dq\right)dt\\
&=& \int\int(t-q)^{2k}|h_n(q)|^2|f(t)|^2 dqdt\\
&=& \int\int(t-q)^{2k}|h_n(q)|^2|f(t)|^2 dtdq,
\eeqat
where lemma \ref{stulpelemma}, the unitarity of the Fourier-Plancherel operator, and Fubini's theorem have been
used. The existence of the last integral implies that $t\mapsto (t-q)^{2k}|f(t)|^2$ is integrable over $\R$ for almost all
$q\in\R$. Thus also $t\mapsto t^{2k}|f(t)|^2$ must be integrable. (In fact, take one $q\in \R$ for which
$t\mapsto (t-q)^{2k}|f(t)|^2$ is integrable and use the fact that there exist positive constants $A, B, M$, such that
$A t^{2k}\leq (t-q)^{2k}\leq B t^{2k}$ for $|t|\geq M$.)
This means that $f$ belongs to the domain of the $k$-th
power of the position operator in $\LkR$ and hence $\vp=Uf\in D(Q^k)$. Conversely, if $\vp=Uf\in D(Q^k)$, the functions 
$t\mapsto |t^{l}||f(t)|^2$ and $q\mapsto |q^l||h_n(q)|^2$ are integrable over $\R$ for all $l \leq 2k$ and hence
$(t,q)\mapsto (t-q)^{2k}|h_n(q)|^2|f(t)|^2$ is integrable over $\RII$. The preceding calculation now shows that
$\vp\in \intsqd{x^k}{E^{|n\ket}}$. The equality $\intsqd{x^k}{E^{|n\ket}}=D(Q^k)$ is thus proved.

The result $\intsqd{y^k}{E^{|n\ket}}=D(P^k)$ is obtained in an analogous manner by using
the fact that the position and momentum operators in $\LkR$ are unitarily equivalent via the Fourier-Plancherel operator $F$.
$\Box$

Now we can determine the operators $\LEnx$ and $\LEny$.
\begin{theorem}\label{LEsxsytheorem}
$\LEnx = p_k^{|n\ket}(Q)$ and $\LEny = p_k^{|n\ket}(P)$, where $p_k^{|n\ket}:\R\to\R$ is the polynomial
\bet
p_k^{|n\ket}(t)=\bra n|(t-Q)^k |n\ket=\sum_{l=0}^k\left(\binc kl(-1)^{k-l}\bra n|Q^{k-l}|n\ket\right) t^l.
\eeqt
\end{theorem}
\noindent {\bf Proof. } Since $p_k^{|n\ket}$ is a polynomial of order $k$ and $Q$ is unitarily equivalent to the position operator
in $\LkR$, the
natural domain of the operator $p_k^{|n\ket}(Q)$ (which is the set $D(Q^k)\cap D(Q^{k-1})\cap\ldots D(Q)$) is equal to that of
$Q^k$. Because $Q$ and $P$ are unitarily equivalent, also $D(p_k^{|n\ket}(P))=D(P^k)$. Thus by the preceding lemma, we
have $D(p_k^{|n\ket}(Q))=D(Q^k) = \intsqd{x^k}{E^{|n\ket}}$ and $D(p_k^{|n\ket}(P))=D(P^k) = \intsqd{y^k}{E^{|n\ket}}$.

Let $\vp,  \psi \in \intsqd{x^k}{E^{|n\ket}}\subset \intd{x^k}{E^{|n\ket}}$.
Let $f=U^{-1}\vp$, $g=U^{-1}\psi$. Since the function
\bet
(q,p)\mapsto q^k \bra \psi|W(-q,p)|n\ket\cc{\bra \vp|W(-q,p)|n\ket}
\eeqt
is integrable over $\RII$, we get
\beat
\bra \psi|\LEnx\vp\ket &=& \int_{\RII} q^{k} dE^{|n\ket}_{\psi,\vp}(q,p)\\
&=& \piker \int q^{k}\left(\int\bra \psi|W(-q,p)|n\ket\cc{\bra \vp|W(-q,p)|n\ket} dp\right)dq\\
&=& \int q^{k}\left(\int \cc{F(\cc{h_n}(\cdot-q)g)(p)}F(\cc{h_n}(\cdot-q)f)(p) dp\right)dq\\
&=& \int q^{k}\left(\int \cc{h_n(t-q)}\cc{g(t)} h_n(t-q)f(t)dt\right)dq\\
&=& \int\left(\int q^{k}|h_n(t-q)|^2 dq\right)\cc{g(t)}f(t)dt\\
&=& \int\left(\int(t-q)^{k}|h_n(q)|^2 dq\right) \cc{g(t)}f(t)dt,\\
&=& \int\ \bra n| (t-Q)^k |n\ket\cc{g(t)}f(t)dt\\
&=& \bra \psi|p_k^{|n\ket}(Q)\vp\ket.
\eeqat
The fifth equality follows from Fubini's theorem, since $(q,t)\mapsto q^{k}|h_n(t-q)|^2\cc{g(t)}f(t)$ is integrable
(because of the inequality
\bet
\left|q^{k}|h_n(t-q)|^2\cc{g(t)}f(t)\right|\leq \frac 12 (1+q^{2k})(|f(t)|^2+|g(t)|^2)|h_n(t-q)|^2
\eeqt
and the proof of lemma \ref{momentdomainlemma}).
The unitarity of $F$ is also used. Since $\psi$ was taken arbitrarily from the dense set $D(Q^k) = \intsqd{f}{E^{|n\ket}}$,
we have $p_k^{|n\ket}(Q)\subset \LEnx$.

The equality $p_k^{|n\ket}(Q) = \LEnx$ follows from the fact that being selfadjoint, the operator $p_k^{|n\ket}(Q)$ cannot
have a proper symmetric extension.

The statement $p_k^{|n\ket}(P) = \LEny$ is obtained in the same manner, since
$p_k^{|n\ket}$ can also be written in the form
$p_k^{|n\ket}(t) = \bra n|(t-P)^k|n\ket$ and $p_k^{|n\ket}(P)$ is selfadjoint.
$\Box$

\

\noindent {\bf Remark.} Since $\bra n |Q^m|n\ket=0$ for odd $m$, and $\bra n |Q^m|n\ket>0$ for even $m$,
only the terms with even $k-l$ are present in the sum defining the polynomial $p_k^{|n\ket}$, and the coefficients
of the corresponding $x^l$ are all strictly positive. 
In particular, $\intL{x^k}{E^{|n\ket}}\neq Q^k$ and $\intL{y^k}{E^{|n\ket}}\neq P^k$ for $k>1$ and
$n\geq 0$, reflecting the difference from the Weyl quantization (\cite[p. 229]{Dubin}), as well as the fact that the Cartesian
margins of $E^{|n\ket}$ are not the spectral measures of $Q$ and $P$.

\

Using Theorem \ref{LEsxsytheorem}, all the operators $\intL{x^k}{E^{|n\ket}}$ and
$\intL{x^k}{E^{|n\ket}}$ can be written in terms of $Q$ and $P$, respectively. In particular, the first and second
moment operators are the following:
\bea
\intL{x}{E^{|n\ket}} &=& Q\nonumber\\
\intL{y}{E^{|n\ket}} &=& P\nonumber\\
\intL{x^2}{E^{|n\ket}} &=& (n+\frac 12)I+Q^2\nonumber\\
\intL{y^2}{E^{|n\ket}} &=& (n+\frac 12)I+P^2. \label{moments}
\eeqa
For the special case of $n=0$, these results were already obtained by formal computations in \cite[pp. 28-29]{Ali}
(without addressing the question on the domains of the operators).
A related result of \cite[p. 140]{Landsman}, however, seems to lack a constant term.

\subsection{The operators $\intL{x^k}{E^T}$ and $\intL{y^k}{E^T}$ with $T=\sum_n w_n |n\ket\bra n|$}

In the next theorem, we need to consider the expressions $\bra n|Q^{2k}|n\ket$. These integrals are
well known and can be calculated e.g. as instructed in \cite[p. 60]{Merzbacher}. We need here only the following fact:
\begin{lemma}
For $n\geq k$, the expression $\bra n |Q^{2k}| n\ket$ can be written as a polynomial in $n$ of order $k$.
\end{lemma}
\noindent {\bf Proof. }
Expressing $Q$ in terms of $A_+$ and $A_-$, we get $\bra n |Q^{2k}| n\ket = \frac {1}{2^k} \|(A_++A_-)^k|n\ket \|^2$.
Because $A_-A_+=N+I$, we can write
\bet
(A_++A_-)^k|n\ket= \sum_{m=0}^{k} a_m|n+k-2m\ket,
\eeqt
where $a_m|n+k-2m\ket = A_+^{k-2m}q^+_m(N)|n\ket$ for $0\leq m\leq \frac k2$ and
$a_m|n+k-2m\ket = A_-^{2m-k}q^-_m(N)|n\ket$ for $\frac k2 \leq m\leq k$, where $q_m^\pm$ are some polynomials with
$2 \text{deg} (q_m^\pm) \pm (k-2m)\leq k$.
Now $a_m^2 = (n+1)(n+2)\ldots (n+k-2m)q^+_m(n)^2$ for $0\leq m\leq \frac k2$ and 
$a_m^2 = n(n-1)\ldots (n-(2m-k)+1)q^-_m(n)^2$ for $\frac k2 \leq m\leq k$, so that each $a_m^2$ is a polynomial in $n$
of order at most $k$, with the coefficient of the highest power positive. Since $a_0^2 = (n+1)(n+2)\ldots (n+k)$, we see
that $\bra n |Q^{2k}| n\ket = \frac {1}{2^k}\sum_{m=0}^k a_m^2$ is a polynomial in $n$ of order exactly $k$.
$\Box$

\begin{theorem}
Let $T=\sum_n w_n |n\ket\bra n|$ be a mixture of the number states and $k\in\N$. Let $p_k^{|n\ket}$ denote the
polynomials defined in Theorem \ref{LEsxsytheorem}, and define
\bet
s_{kl} = \binc {k}{l}\sum_{n=0}^\infty w_n \bra n|Q^{k-l}|n\ket \ (\leq \infty)
\eeqt
for $0\leq l\leq k$.
\begin{itemize}
\item[(a)] $\intsqd{x^k}{E^T} \neq \{0\}$ if and only if
\be\label{nseries}
\sum_n n^k w_n <\infty,
\eeq
in which case $s_{kl}<\infty$ for all $0\leq l\leq k$, $\intd{x^k}{E^T}= \intsqd{x^k}{E^T} = D(Q^k)$, and  
\bet
\intL{x^k}{E^T} = \sum_{l=0}^k s_{kl} Q^l.
\eeqt
In particular, the operator $\intL{x^k}{E^T}$ is then selfadjoint.
\item[(b)] The statement (a) holds true when ''$x$'' and ''$Q$'' are replaced by ''$y$'' and ''$P$''.
\end{itemize}
\end{theorem}
\noindent {\bf Proof. } Let $k\in\N$ be fixed. According to Proposition \ref{LsubLnlemma}, Lemma \ref{momentdomainlemma} and
Theorem \ref{LEsxsytheorem} we have,
\be\label{LsubLn}
\intL{x^k}{E^T}|_{\intsqd{x^k}{E^T}} \subset  \sum_n w_n \intL{x^k}{E^{|n\ket}} = \sum_n w_n p_k^{|n\ket}(Q)
\eeq
(where the series of operators are understood in the same sense as in Proposition \ref{LsubLnlemma}).
Let $\vp\in D(Q^k)\supset D(\sum_n w_n p_k^{|n\ket}(Q))$ and $\vp\neq 0$.
Let $A_{\vp,\vp}^{|n\ket}$ be the density function of the positive measure $E_{\vp,\vp}^{|n\ket}$.
By the definition of the square integrability
domain, $\vp\in\intsqd{x^k}{E^T}$ if and only if the function $x^{2k}\sum_n w_n A_{\vp,\vp}^{|n\ket}$ is integrable.
By the monotone convergence theorem, the latter statement is equivalent to
\be\label{vpseries}
\sum_n w_n \int x^{2k} dE_{\vp,\vp}^{|n\ket} <\infty.
\eeq
According to the proof of Lemma \ref{momentdomainlemma},
\beat
\int x^{2k} dE_{\vp,\vp}^{|n\ket} &=& \int\int(t-q)^{2k}|h_n(q)|^2|(U^{-1}\vp)(t)|^2 dtdq\\
&=& \sum_{l=0}^{2k} \binc {2k}{l} (-1)^{2k-l}\left(\int t^l|(U^{-1}\vp)(t)|^2dt\right)\bra n | Q^{2k-l}|n\ket.
\eeqat
Since $\bra n |Q^m |n\ket= 0$ for odd $m\in\N$, only the terms with even $l$ are present in the above sum. Because
$U^{-1}\vp\neq 0$, these terms are all strictly positive.
Since $0 < \bra n |Q^{2l} |n\ket \leq \int (1+q^{2k})|h_n(q)|^2dq = 1+\bra n | Q^{2k} |n\ket$ for all $l\leq k$ and
$\sum_n w_n=1$, the convergence of the series $\sum_n w_n \bra n | Q^{2k} |n\ket$ implies the
convergence of each series $\sum_n w_n \bra n | Q^{2l} |n\ket$ and $s_{kl}$, $l\leq k$.
Thus, it follows that a nonzero vector in $D(Q^k)$ belongs to $\intsqd{x^k}{E^T}$ if and only if the series
\be\label{Qseries}
\sum_n w_n  \bra n |Q^{2k}| n\ket
\eeq
converges. By the preceding lemma, this is equivalent to (\ref{nseries}). Since $\intsqd{x^k}{E^T}\subset D(Q^k)$ by
(\ref{LsubLn}), we have
shown that $\intsqd{x^k}{E^T}\neq \{0\}$ if and only if (\ref{nseries}) holds, and in that case,
$\intsqd{x^k}{E^T} = D(Q^k)$ and $s_{kl}<\infty$ for $l\leq k$.

From the definition of the polynomials $p_k^{|n\ket}$ we see that if (\ref{Qseries}) converges, then (using (\ref{LsubLn}))
we get 
\bet
\bra \psi |\intL{x^k}{E^T}\vp\ket = \sum_n w_n \bra \psi |p_k^{|n\ket}(Q)\vp\ket = \sum_{l=0}^{k}  s_{kl} \bra \psi | Q^l\vp\ket 
\eeqt
for each $\vp\in D(Q^k)$ and $\psi\in\hil$.
Thus, $\sum_{l=0}^k s_{kl} Q^l\subset \intL{x^k}{E^T}$ (now $D(\sum_{l=0}^k s_{kl} Q^l)= D(Q^k)$, because $s_{kk}=1$).
Since the operator $\sum_{l=0}^k s_{kl} Q^l$ is selfadjoint and $\intL{x^k}{E^T}$ is symmetric,
the statement (a) has been proved.

The result (b) is obtained in the same manner, since
\bet
\intL{y^k}{E^T}|_{\intsqd{y^k}{E^T}} \subset  \sum_n w_n \intL{y^k}{E^{|n\ket}} = \sum_n w_n p_k^{|n\ket}(P),
\eeqt
$\bra n |Q^m |n\ket = \bra n |P^m |n\ket$ for all $m, n \geq 0$ and
\beat
\int y^{2k} dE_{\vp,\vp}^{|n\ket} &=& \int\int(t-p)^{2k}|Fh_n(p)|^2|(FU^{-1}\vp)(t)|^2 dtdp \\
&=& \sum_{l=0}^{2k} \binc {2k}{l} (-1)^{2k-l}\left(\int t^l|(FU^{-1}\vp)(t)|^2dt\right)\bra n | P^{2k-l}|n\ket
\eeqat
for all $\vp\in D(P^k)$.
$\Box$

\

\noindent {\bf Remark.} For each $k\in\N$, there are states $T$ of the form (\ref{mixed}) such that (\ref{nseries})
does not converge (so that $\intsqd{x^k}{E^T}=\{0\}$), but all the series $s_{kl}$ do. In that case,
$\sum_{l=0}^ks_{kl} Q^l$ is still a well-defined selfadjoint operator with the domain $D(Q^k)$.
An example of such a state for $k\in \N$ is given by $t_n = \frac {1}{S n^{k+1}}$, where $S=\sum_n n^{-(k+1)}$.
We do not know whether there are any nonzero vectors in the domain $\intd{x^k}{E^T}$ then.

\section{Operator integrals of some polynomials}

In this last section, we use Theorem \ref{LEsxsytheorem} to determine the operator integrals for certain types of
polynomials. To that end, let $h, h_1, h_2$ be real polynomials defined by $h(t)=\sum_{l=0}^k a_l t^l$, $a_k\neq 0$, and
$h_i(t) = \sum_{l=0}^{k_i} a_{i,l} t^l$, $a_{i,k_i}\neq 0$, $i=1,2$.

\

\noindent {\bf The operators $\intL{h\circ x}{E^{|n\ket}}$ and $\intL{h\circ y}{E^{|n\ket}}$.}
Let $\psi,\vp\in\hil$.
There exist positive constants $M, A, B$ such that $A |t^k|\leq |h(t)|\leq B|t^k|$ for $|t|\geq M$, which implies that
the function $h\circ x$ (i.e. $(q,p)\mapsto h(q)$) is $E^{|n\ket}_{\psi,\vp}$-integrable if and only if $x^k$ is such, and in that
case, $\int h\circ x \ dE^{|n\ket}_{\psi,\vp} = \sum_{l=0}^k a_l \int x^l dE^{|n\ket}_{\psi, \vp}$. 
Since $\intd{x^k}{E^{|n\ket}}=D(Q^k)\subset D(Q^l)=\intd{x^l}{E^{|n\ket}}$ for $l\leq k$, it follows that
$\intd{h\circ x}{E^{|n\ket}} = D(Q^k)$, and
\be\label{intpx}
\intL{h\circ x}{E^{|n\ket}}= \sum_{l=0}^k a_l p_l^{|n\ket}(Q).
\eeq

Naturally, a similar result holds for the function $h\circ y$.

\

\noindent {\bf The operators $\intL{h_1\circ x+ih_2\circ y}{E^{|n\ket}}$.}
Let $\psi,\vp\in\hil$. The function $h_1\circ x+ih_2\circ y$ is $E^{|n\ket}_{\psi,\vp}$-integrable if and only
if both $h_1\circ x$ and $h_2\circ y$ are such, and in that case
$\int (h_1\circ x+ih_2\circ y) \ dE^{|n\ket}_{\psi,\vp}=\int h_1\circ x \ dE^{|n\ket}_{\psi,\vp}+i\int h_2\circ y \ dE^{|n\ket}_{\psi,\vp}$.
Thus, we have $\intd{h_1\circ x+ih_2\circ y}{E^{|n\ket}} = D(Q^{k_1})\cap D(P^{k_2})$, and
\be\label{intpi}
\intL{h_1\circ x+ih_2\circ y}{E^{|n\ket}}=\sum_{l=0}^{k_1} a_{1,l}p_l^{|n\ket}(Q) +
i \sum_{l=0}^{k_2} a_{2,l} p_l^{|n\ket}(P).
\eeq

\

\noindent {\bf The operators $\intL{h_1\circ x+h_2\circ y}{E^{|n\ket}}$ with $k_i$ even and $a_{i,k_i}>0$.}
Assume that $k_i$ is even, and $a_{i,k_i}>0$, $i=1,2$. Then we can choose positive constants
$M, A_i, B_i$ such that $h_i(t)\geq 0$ and $A_i t^{k_i} \leq h_i(t) \leq B_i t^{k_i}$ for
$|t|\geq M$ and $i=1,2$. This implies that the function $h_1\circ x+h_2\circ y$ is $E^{|n\ket}_{\psi,\vp}$-integrable
for $\psi, \vp\in\hil$ if and only if both $x^{k_1}$ and $y^{k_2}$ are such. We get
$\intd{h_1\circ x+h_2\circ y}{E^{|n\ket}} = D(Q^{k_1})\cap D(P^{k_2})$, and
\be\label{intpit}
\intL{h_1\circ x+h_2\circ y}{E^{|n\ket}}=\sum_{l=0}^{k_1} a_{1,l} p_l^{|n\ket}(Q) +
\sum_{l=0}^{k_2} a_{2,l} p_l^{|n\ket}(P).
\eeq

\

\noindent {\bf A note on the operators $\intL{ x \pm i y}{E^{|n\ket}}$ and $\intL{x^2+y^2}{E^{|n\ket}}$.}
The above observations show that, in particular, $\intL{ \qker (x \pm i y)}{E^{|n\ket}} = \qker (Q\pm iP)$, and
$\intL{\frac 12 (x^2+y^2)}{E^{|n\ket}} = \frac 12 (Q^2 +P^2) +(n+\frac 12)I$. These operator integrals
have already been
determined in \cite{Lahti} (using a different method) to be the following: $\intL{ \qker (x \pm i y)}{E^{|n\ket}} = A_{\mp}$ and
$\intL{\frac 12 (x^2+y^2)}{E^{|n\ket}} = N+(n+1)I$. The fundamental operator equalities
(\ref{AQP}) and (\ref{numberop}) now show that the results are indeed consistent.

\

\noindent {\bf Acknowledgement.} The authors thank Drs. Daniel Dubin and Mark Hennings for pointing out some
details concerning the validity of the operator equality (\ref{numberop}).

\end{document}